\documentclass[aps,twocolumn]{revtex4}

\begin{document}

\title{The Future of Computation%
\footnote{Invited lecture at the Workshop on Quantum Information,
Computation and Communication (QICC-2005), IIT Kharagpur, India,
February 2005, {\tt quant-ph/0503068}.}
}
\author{Apoorva Patel}
\affiliation{\rm Centre for High Energy Physics,\\
Indian Institute of Science, Bangalore-560012, India\\
E-mail: adpatel@cts.iisc.ernet.in}

\begin{abstract}
``The purpose of life is to obtain knowledge, use it to live with as much
satisfaction as possible, and pass it on with improvements and modifications
to the next generation.''
This may sound philosophical, and the interpretation of words may be
subjective, yet it is fairly clear that this is what all living
organisms---from bacteria to human beings---do in their life time.
Indeed, this can be adopted as the information theoretic definition of life.
Over billions of years, biological evolution has experimented
with a wide range of physical systems
for acquiring, processing and communicating information.
We are now in a position to make the principles
behind these systems mathematically precise,
and then extend them as far as laws of physics permit.
Therein lies the future of computation, of ourselves, and of life.
\end{abstract}
\maketitle

\section{Computation}

The silicon transistor was invented about half a century ago.
Since then the semiconductor technology has grown at a rapid pace to
pervade almost all aspects of our lives.
This growth has been so explosive---doubling the number of transistors
on a chip every 18-24 months according to Moore's law---that many choices
made in constructing the theoretical framework of computer science
(see for example, Ref.\cite{neumann}) were almost forgotten.
Computer architecture became essentially synonymous with digital electronic
circuits implementing Boolean operations, pushing aside other competing models.
Developments in quantum computation during the past decade have led us to
question this attitude, and brought in focus the fact that there is much
more to information theory than just Boolean logic.
The concept of what is computable and what is not has not changed,
but the criteria determining how efficiently a computational task
can be implemented have been altered.
The reason behind this change is that some of the implicit assumptions
of theoretical computer science are too restrictive,
when compared to physically realisable models.
Computational power of a framework can be enhanced by discarding such
unnecessary assumptions.
Is there then a framework which would allow one to design the optimal
computer for a given task?
My aim in this article is to systematically formulate a generalised
information theory, based on many physical examples we encounter in
the world around us, that can cover all types of computational schemes.

Information theory deals with two broad areas, communication and computation.
Communication is quite simple---the receiver gets
whatever is sent by the sender.
There is no processing in the middle, unless there is some unwanted noise.
Computation is more complex---the input is intentionally manipulated
by an external agency to produce an output.
Though the steps involved in manipulating information can be expressed in
abstract mathematical terms, they have to be implemented by physical devices.
We therefore must address the role of physical laws in the construction
of a general information processing framework.

\begin{figure}
\begin{picture}(200,100)(0,0)
\put(60,28){\framebox(60,40){\shortstack{Physical\\ Device}}}
\put(0,45){Input}
\put(30,48){\vector(1,0){30}}
\put(160,45){Output}
\put(120,48){\vector(1,0){30}}
\put(70,90){Instructions}
\put(90,88){\vector(0,-1){20}}
\put(50,0){Oracles/Look-up Tables}
\put(90,24){\vector(0,1){4}}
\multiput(90,8)(0,4){4}{\line(0,1){2}}
\end{picture}
\vspace{2mm}

\caption{Schematic representation of a computer.\\
Every computation may not use oracles or look-up tables.}
\end{figure}
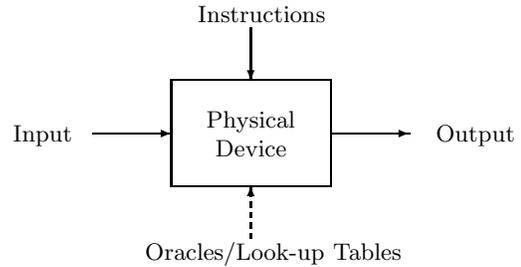

\section{The Importance of Physics}

A schematic representation of a computer is shown in Fig. 1,
where I have deliberately replaced the standard input and output
slots by data and knowledge respectively.
This is not a question of merely adding an appropriate interface.
Rather it restricts, as we will see, of what can be computed in
the real world, and of what use that may be.

Data list physical properties of a system.
They describe a particular realisation of the physical system,
amongst its many possible states.
Data are often obtained by experimental observations of the system,
and generally provide the starting point of a computational process.
I stress that data are always firmly rooted in physical characteristics,
and should not be separated from them.

Information is the abstract mathematical property
obtained by detaching all the physical characteristics from data.
It just becomes a measure of the number of possible states
of the system \cite{shannon}.
This mathematical abstraction proves to be very useful,
because in dealing with information, at no stage one has to worry about
where the information came from or what it means.
The physical realisation of information may change according to
the convenience of the task to be carried out.
For example, our electronic computers compute using electrical signals,
but store the results on the disk using magnetic signals;
the former realisation is suitable for fast processing,
while the latter is suitable for long term storage.
Many common computational tasks manipulate information
without bothering about its meaning,
e.g. compress data, quantify error rate, devise codes, and so on.

While extraction of abstract information from data allows one to
formulate precise mathematical rules for its systematic analysis,
the abstraction also brings in a limitation.
In order to implement mathematical algorithms,
one must map information to physical properties,
and the types of manipulations that can be carried out
are limited by the types of physical devices available.
For example, we use various programming languages to implement
mathematical algorithms on a computer.
On the other hand, the electronic computer hardware responds
only to voltages and currents.
So a whole hierarchy of translation machinery is constructed,
involving compilers and operating systems,
to convert the algorithms to binary machine codes
and then map them on to off/on states of silicon transistors.
An important consequence of this physical dependence is that
the efficiency of a computational task cannot be determined solely by
its mathematical algorithm---the efficiency depends on the algorithm as
well as on the properties of the physical device that implements it.

The role of physical properties is also inevitable in adding
a sense of purpose to information, and converting it to knowledge.
If the receiver does not understand the language of the message,
he will just have random looking symbols and no meaningful interpretation.
He will gain the knowledge contained in the message,
only when he figures out the language.
(As a matter of fact, the whole subject of cryptography
is based on sending the information but concealing its language.)
A common language can be established between the
sender and the receiver only by physical means.
Of course, once a common language is established,
it can be used repeatedly in an abstract manner.
In case of the most primitive (or low level) messages,
there is no luxury of abstract languages---the only language that exists
is the one labeled by physical properties.
In such cases, the physical objects that carry the message have to convey
the information as well as its interpretation to the receiver.
Once again, an optimal language can only be designed
if the available physical means are known.

To summarise, \emph{data is not information and information is not knowledge}.
We have instead,
\begin{center}
Information = Data - Physical Realisation ,\\
Knowledge = Information + Interpretation .
\end{center}
Abstract information theory does not tell us
what physical realisation would be appropriate for a particular message,
nor does it tell us the best way of implementing a computational task.
To make such choices, we must look at the physical resources available,
i.e. analyse the type of information and not the amount of information.

\section{The Importance of Biology}

The number of fundamental physical interactions is rather small,
and that limits the possible physical realisations of a computer.
A variety of computational schemes can still be created, however,
by combining the fundamental ingredients in different ways.
To get an idea about the multitude of physical resources that can be used
to process information, it is instructive to look at biological systems.
Over billions of years, evolution has had plenty of time (which we do not
and cannot have) for experimentation with a wide range of physical systems,
and selected the best of the available options.

We are accustomed to looking at our computers from the top level
down---from the abstract mathematical operations to the transistors
embedded in silicon chips.
On the other hand, to be able to design efficient computers,
we must study them from the bottom level up---from
the elementary building blocks to the complicated languages.
Biological systems have indeed evolved in this manner,
from biomolecular interactions to multicellular organisms,
and have explored a variety of options along the way.
But biological evolution has been rather blind.
Mutations first construct different combinations
of elementary building blocks,
and then decide whether or not the constructions would be of any use.
By a careful analysis of the evolutionary patterns, were we to figure
out the underlying principles of what can be useful in what context,
then we can be much quicker than biological evolution in finding ways
of efficient information processing.

\begin{table}[b]
\begin{center}
\begin{tabular}{lll}
Organism      & Messages             & Physical Means  \\
\hline
Single cell   & Molecular            & Chemical bonds, \\
              & (DNA, Proteins)      & Diffusion       \\
\hline
Multicellular & Electrochemical      & Convection,     \\
              & (Nervous system)     & Conduction      \\
\hline
Families,     & Imitation, Teaching, & Light, Sound    \\
Societies     & Languages            &                 \\
\hline
Humans        & Books, Computers,    & Storage devices,\\
              & Telecommunication    & Electromagnetic \\
              &                      & waves           \\
\hline
Cyborgs ?     & Databases            & Merger of brain \\
              &                      & and computer    \\
\end{tabular}
\caption{Different levels of biological information systems.}
\end{center}
\end{table}

The multitude of information systems discovered by the living organisms
in the course of evolution. together with their physical implementations,
are presented in Table I.
It is fairly obvious that evolution has progressively discovered
higher levels of communication mechanisms, whereby\\
$\bullet$ Communication range has expanded,\\
$\bullet$ Physical contact has reduced,\\
$\bullet$ Abstraction has increased,\\
$\bullet$ Succinct language forms have arisen,\\
$\bullet$ Complex translation machinery has developed.\\
A striking feature of evolution is that the discovery of
new physical means of processing information has led to new modes of
information storage and task execution
(e.g. memory has progressed from DNA to brain to books and databases),
and the interlinking of these modes has proved to be highly advantageous.
With a bit of hindsight, we can say that knowledge has proved to be
the driving force behind evolution.
At the same time, the power of knowledge provides only the direction
for evolution, and not the destination.
How far one can proceed along a given direction is ultimately dictated
by laws of physics.

Our computers were designed to mimic operations of the brain \cite{neumann}.
A hierarchical pattern of operations, similar to living organisms,
is built in to them, as illustrated in Table II.
The present day computers are largely based on, and therefore also limited by,
the information processing that can be carried out using electrical circuits.
We must look beyond---to other physical realisations---to expand their
capabilities.

\begin{table}[b]
\begin{center}
\begin{tabular}{lll}
Living organisms         &  Task       & Computers         \\
\hline
Signals from environment & Input       & Data              \\
Sense organs             & High level  & Pre-processor     \\
Nervous system           & Translation & Operating system  \\
and Brain                &             & and Compiler      \\
Electrochemical signals  & Low level   & Machine code      \\
Proteins                 & Execution   & Electrical signals\\
DNA                      & Programme   & Programmer        \\
\end{tabular}
\caption{Comparison of hierarchical processing of information in
living organisms and in computers.}
\end{center}
\end{table}

\section{Physical Criteria}

To expand the scope of information theory,
I generalise the notion of a message from \emph{a sequence of letters}
to \emph{a collection of building blocks}.
Collections can be labeled according to the number of external
space-time dimensions where the building blocks are arranged.
The building blocks themselves can be characterised by their properties,
external as well as internal.
As already emphasised, the appropriate building blocks and collections
for a given task have to be selected based on physical principles.
Furthermore, the selection can be optimised depending on
what is available and what is to be accomplished.
The most efficient computers are those that
reliably accomplish their tasks using the least amount of resources.
The optimisation process is thus guided by two principles:
(i) minimisation of errors, and (ii) minimisation of physical resources.
These principles often impose conflicting demands,
and one has to learn how to tackle them in the process of computer design.
Let us analyse them in turn.

Laws of thermodynamics imply that unwanted disturbances can never be
completely eliminated---errors are an unavoidable fact of life.
So we must develop strategies to keep the error rate in control.
The system can be protected from external disturbances by shielding.
On the other hand, the system can be guarded against internal fluctuations
only if the information processing language is based on discrete variables
(as opposed to continuous variables).
Allowed values of fundamental physical variables are often continuous,
in which case a set of non-overlapping neighbourhoods
of discrete values can be chosen as the discrete variables.
The advantage is that the discrete variables remain unaffected,
even when the underlying continuous variables drift,
as long as the drifts keep the values within the assigned neighbourhoods.
This is the common procedure of \emph{digitisation}, it eliminates small
fluctuations and leads to the framework of \emph{bounded error computation}.
For example, my handwriting is not the same as yours,
nor is my accent the same as yours.
Yet you can figure out what I write or what I speak,
because the letters and sounds of our languages are discrete.
A close match---and not an exact match---is sufficient for you
to understand what I convey.

In a language based on continuous variables, it is not possible to tell
apart what is unwanted noise and what is a genuine transformation.
On the other hand, in a digitised language, all small fluctuations are
interpreted as unwanted noise, and are eliminated by resetting the
variables to their discrete values once in a while.
All large changes are interpreted as genuine transformations,
and so large erroneous changes still persist in a digitised language.
Digitisation is thus worthwhile, when large erroneous changes are rare.
In fact, large erroneous changes can be eliminated too,
provided their rate falls below a certain threshold,
with the help of error correcting codes (based on redundancy and nesting).

It is useful to note that quantum physics at the atomic scale
automatically provides discrete variables,
e.g. finite size of atoms leads to lattices,
and discrete energy levels lead to characteristic transitions.
In other cases, there is a loss of precision when changing from
continuous variables to discrete ones, e.g. discrete variables can
produce integers and rational numbers but not irrational numbers.
Yet the framework of bounded error calculations is immensely useful,
because in all our practical applications we never need results with
infinite precision; as long as results can be obtained within a
a prespecified non-zero tolerance limit, they are acceptable.
The error rate depends on the physical device processing information,
and the tolerance limit is specified by the computational task to be
carried out---bringing them together is a question of computer design.

Physical resources to be optimised include space, time and energy.
Minimisation of spatial resources means carrying out
the computational task using as few physical components as possible,
e.g. memory and disk space in our digital computers.
In addition to finding a software algorithm which requires the smallest
number of variables, this also requires selecting elementary hardware
components that are simple and easily available.
This is the common choice at the lowest level of information processing,
and complicated systems are then constructed by packing a large number
of components in a small volume.
Moreover, the information content of a language resides in its
aperiodic random patterns; correlations and repetitive structures
in a language waste spatial resources.
The language is therefore most versatile when its building blocks
can be arranged in as many different ways as possible.

Minimisation of temporal resources means finding an algorithm with the
smallest number of execution steps, and also finding hardware components
that allow fast implementation of computational instructions.
Often a trade-off is possible between spatial and temporal resources,
and specific choices are made depending on what is more important,
e.g. parallel computers save on time by using more hardware.

A computer is a driven physical system,
with irreversible operations of resetting and erasure.
So, according to thermodynamical laws,
a source of free energy is required to run it.
This thermodynamical limit is not of much practical relevance, however,
because available physical devices are nowhere near that efficiency.
Energy consumption during information processing depends almost
entirely on the choice of hardware technology.
The best strategy is to make the hardware components as tiny and as cheap
as possible, so that they can carry out their tasks consuming little energy,
and also recycle energy wherever possible.

Now we can see that conflicts arise amongst these optimisation guidelines.
Tiny components and fast operations are less reliable and increase noise,
error correction procedures add overheads to physical resources,
more precise operations demand more energy, segregating different
ingredients of a computational task and assigning them to specialised
components increases the reliability of computation but increases
resource requirements, and so on.
Depending on how much weight is assigned to which criterion,
different languages can be designed to implement the same computational task.
We know by experience that when the languages are versatile enough,
information can be translated from one language into another by replacing
one set of building blocks and operations by another.
Subjective (and historical) choices have often determined specific realisations.

When a number of choices are available, the language with the smallest set
of building blocks has a unique status in the optimisation procedure:\\
(a) Generically, physical hardware properties have a fixed range of values.
Decreasing the number of discrete states allows them to be put
as far apart from each other as possible within that range.
This dispersal minimises misidentification,
and provides the largest tolerance against errors.
For example, silicon transistors are powerful non-linear electrical devices,
but they are used in digital computers only as two saturated extreme states.\\
(b) Reduction of possible physical states of elementary components
simplifies the instruction set needed to manipulate them,
and also the possible types of connections amongst the components.
For example, with our decimal number system, we had to learn $10\times10$
tables to do arithmetic in our primary schools.
With the binary number system, only $2\times2$ tables are needed,
and our computers implement them with the Boolean operations
XOR (for addition) and AND (for multiplication).\\
(c) A small number of discrete states increases the depth of computation,
i.e. the number of building blocks required to represent a fixed amount
of information.
But with only a small number of states and instructions,
elementary components can be made small and individual instructions fast.
Typically, high density of packing and quick operations
more than make up for the increase in the depth of computation,
and the overall requirement for physical resources goes down.\\
(d) At the lowest level of information processing,
translation of languages is not possible, and only a handful of
instructions related to physical responses of the hardware exist.
The simplest language then has a distinct advantage,
and it becomes the universal language for that particular hardware.

\section{Types of Collections}

We are now in a position to look at some examples of information processing
systems, and understand how well they implement the optimisation principles. 
Messages are constructed by linking the basic components---the building
blocks of the language---in a variety of arrangements.
The information contained in a message depends
on the values and positions of the building blocks.
Any language that communicates non-trivial information
must have the flexibility to arrange its building blocks
in different ways to represent different messages,
and such arrangements must involve specific physical phenomena.

Let us look at possible collections of building blocks.\\
$\bullet$ \emph{0-dim:}
Such a collection requires multiple building blocks
to be at the same point in space and time.
This is the phenomenon of superposition,
which is a generic property of waves.
Superposition allows many signals to be combined together,
and then also be manipulated together, but at the end
only one of the signals can be extracted from the collection.
For example, radio and television broadcasts combine
multiple electromagnetic signals together,
and the receiver extracts the desired signal (only one at a time)
by tuning to the corresponding frequency.\\
$\bullet$ \emph{1-dim:}
Here the message is an ordered sequence of building blocks.
It is the most common form used in conventional information theory.
Mathematically, the collection is expressed as a tensor product
of individual components.
The ordering of the sequence can be either in space or time,
e.g. our written and spoken languages.\\
$\bullet$ \emph{2-dim:}
Higher than one dimensional collections can be viewed
as combinations of multiple ordered sequences.
A common situation is that of parallel computation,
based on multiple similar information processing units.
Such parallelism allows an unusual feature,
namely information can reside in correlations amongst sequences
without being present in any individual sequence.
Biological systems have efficiently exploited this feature,
whereby gradients are detected at the cost of redundancy.
The most common implementation uses multiple detectors
(e.g. eyes and ears) to estimate direction and distance,
either by parallax removal or by detecting concentration gradients.
Such systems have been left out of our computers---our computers
are not at all efficient at finding gradients.
We are gradually learning to use such systems for certain tasks,
e.g. very long base-line interferometry (VLBI) in astronomy,
global positioning system (GPS) in geography,
and space-time codes in electronic communications.\\
$\bullet$ \emph{3-dim}:
Such collections describe the physical structure of an object
in our three dimensional space.
Structural information is useful for establishing lock-and-key mechanisms,
which can trigger an appropriate response.
For example, proteins and other biomolecules use such a system,
based on extremely short-ranged molecular forces,
to carry out various tasks in living organisms.\\
$\bullet$ \emph{4-dim}:
This would be a complete description of any event, either past or future,
in our universe with one time and three space dimensions.
Such a description would contain all the information about a system,
that can ever be extracted.
On the other hand, it is too much for our common use,
and we typically use only a smaller dimensional subsystem for our tasks.

It is not necessary that a collection of building blocks
be restricted to a fixed dimensionality.
In fact, computational capability of a system can be vastly enhanced
by combining features of different dimensionalities.
For example, the current framework of quantum computation \cite{nielsen}
uses collections of both zero and one dimensions.
The phenomenon of superposition, combined with the conventional sequence of
qubits, leads to the unusual possibility of quantum entanglement of states.
It is this combination which enables a quantum computer to solve certain
problems much more efficiently compared to a classical computer.

Another example of multiple dimensionality is provided by proteins,
which combine features of both one and three dimensional collections.
The one dimensional form of proteins is convenient
for efficient synthesis through polymerisation of amino acids,
and also for crossing cellular membranes through narrow channels.
The three dimensional form is suitable for carrying out various functions
through highly selective binding to other molecules of complementary shape.
The mechanism for realising both these forms is based on the property
that proteins are physical systems poised at the edge of criticality.
Small changes in suitable external parameters
(e.g. concentration of a denaturant or pH of the solvent)
can unfold the protein to its polypeptide chain form,
or conversely, fold it into its three dimensional native form.

Superposition, parallax, phase transitions,
are all well understood physical phenomena.
The examples above illustrate how the capability of an information
processing system can be enhanced by incorporating them in physical devices.
Our conventional framework of computation has barely made a start in that
direction.

\section{Types of Building Blocks}

Physical properties of building blocks, in both internal and external space,
are generically organised in terms of groups.
(There is an implicit assumption here that
we can recognise the same object in different manifestations,
just as we can identify the same person wearing different clothes.)
As discussed above, for a given information processing task,
the smallest discrete group that can implement it
is the ideal candidate for the optimal language.
Often the group of physical properties is a continuous one,
then we must look for its smallest yet faithful discrete subgroup.
We have also observed earlier that because of unavoidable noise,
a discrete building block of a language is associated not with just
a point on the group manifold but with a neighbourhood of the point.
Thus to specify the building blocks completely,
we have to describe neighbourhoods of discrete group elements.

The algebra of any group is fully specified in terms of its generators.
The number of independent generators gives the dimensionality of the group.
In case of continuous groups, the generators define a vector space.
In a $d$-dimensional group manifold,
any group element is specified by $d$ coordinates,
and one more parameter is needed to specify the size of its neighbourhood.
For generic manifolds, such a $(d+1)$ parameter object is called a simplex.
It provides the simplest specification of an elemental group volume
which faithfully realises all group properties.
The smallest discrete realisation of any group, therefore,
corresponds to replacing the entire group by a single simplex.

Sometimes the dual (Fourier) space of representations provides a more convenient
description of the group than the coordinates specifying the group elements.
In that case, the minimal set of $(d+1)$ elementary building blocks
is formed by the $d$-dimensional fundamental representation
and the $1$-dimensional identity representation.
Any other representation of the group can be obtained by putting together
several of these simplest representations.

In general, the building blocks are completely specified in terms of
two discrete groups, one for the external properties and one for the
internal ones (one of the groups may be trivial in some cases).
Let us look at the minimal set of building blocks for some common groups:\\
$\bullet$ \emph{1-dim:}
Groups with a single generator include cyclic groups,
the set of integers and the real line.
The minimal simplex in this case has just two points, $Z_2=\{0,1\}$.
It forms the basis of Boolean arithmetic widely used in digital computers.
The binary language can be easily extended to a $d$-dimensional situation,
as the Cartesian product $(Z_2)^d$, and is therefore convenient
as a general purpose language in handling a variety of problems.\\
$\bullet$ \emph{2-dim:}
The simplex for two dimensional geometry is a triangle.
Triangulation is useful in discrete description of arbitrary surfaces.
At nanoscale, its dual hexagonal form can be realised in terms of
the $sp^2$-hybridised orbital structure of graphite sheets,
which may become useful in lithographic techniques.\\
$\bullet$ \emph{3-dim:}
In three dimensional space, the simplex is a tetrahedron.
At molecular level, $sp^3$-hybridised orbitals provide its dual form.
Arbitrary structures can be created by gluing tetrahedra together.
Tetrahedral geometry based on properties of carbon provides a convenient
way to understand the three dimensional language of proteins \cite{carbon}.\\
$\bullet$ $SU(2):$
This is also a group with three generators,
up on which the description of quantum bits is based.
Arbitrary states of a qubit, including the mixed states arising from
decoherence (i.e. environmental noise), can be fully described using
a density matrix, which is a linear combination of four operators
$\{1,\sigma_x,\sigma_y,\sigma_z\}$.

There are certainly many other discrete groups of physical properties,
e.g. permutation, braid and symplectic groups.
Also, several large groups have been used in error correcting codes
and cryptography, but not for processing information.
Whether such groups can be used for constructing a computational
framework or not is an exciting question to be explored.

\section{Types of Processing}

Once the physical properties of the building blocks are selected,
i.e. the discrete groups describing their external and internal properties, 
the possible computational operations are just group transformations.
Different physical means are needed in case of different groups,
and what is possible and what is not depends on the available technology.
It is straightforward to list the possibilities:\\
$\bullet$ \emph{0-dim:}
The only mathematical operation allowed with superposition is addition.
Addition is commutative,
and interference effects produced by it are physically useful.\\
$\bullet$ \emph{1-dim:}
This is the most common realisation, where two different
group operations of addition and multiplication are possible.
Both operations are commutative, their combination obeys
a distributive rule, and all our arithmetic is based on them.
In mathematical terms, $Z_2$ is a field---the smallest one.\\
$\bullet$ \emph{dim$>$1:}
In higher dimensions, addition generalises to translation.
The obvious generalisation of multiplication is scale transformation,
but the scope of multiplication can be expanded to include rotations
as well (which can be viewed as multiplication by a matrix).
Rotations are commutative in two dimensions, but non-commutative for $d>2$.
Discrete operations of translation, rotation and scale transformation
can be realised on a lattice made of simplicial building blocks.
The algebra generated by them is much more powerful than common arithmetic.

Clearly, more and more group operations become possible
as the dimensionality of the group increases.
Direct physical implementation of a complicated group operation
can substantially reduce the depth of computation.
For example, steps of a quantum algorithm can be represented
in classical language as multiplication of unitary matrices
with superposed state vectors.
Such a multiplication is a single operation on a quantum computer,
but an elaborate procedure on a classical computer,
and therein lies the physical advantage of a quantum computer.
From this point of view, we have hardly begun to explore
the power of non-commutative group algebra.

\section{The Future}

We have seen that the scope of information processing can be vastly
enhanced by looking at a message as a collection of building blocks.
Physical optimisation criteria require discrete languages, versatile
operations, special purpose components and tiny building blocks.
But beyond that, there is lot of freedom in the choice of building blocks.
A variety of computational frameworks can be constructed by appropriate
choice of (i) the dimensionality for the arrangement of building blocks,
and (ii) the group structure for the properties of building blocks.
I have described several physical computational systems above,
and pointed out the choices inherent in their design.
It is natural to look for other possible choices, which may help
in finding the optimal hardware design for a given computational task,
and which may lead to novel computational schemes:\\
$\bullet$ Operations of calculus, such as differentiation and integration,
are easier to carry out using continuous variables instead of discrete ones.
Although digitisation is necessary to control errors,
it does not have to be imposed at every computational step.
So the framework of analogue computation,
occasionally punctuated by digitisation,
may turn out to be convenient for implementing operations of calculus.\\
$\bullet$ Correlations between parallel streams of data can convey
information that is not at all easy to convey using a single stream.
Quantum entanglement and parallax removal offer unique opportunities
in such systems.
Mobile telephone companies are exploring space-time communication codes
that use multiple transmitters and antennae.
Such codes can increase bandwidth as well as overcome noise
(as exploited in astronomical detection of highly faint objects).\\
$\bullet$ A fractal arrangement would be an unusual collection
of building blocks.
Such self-similar patterns occur in concatenated error correcting codes,
but can they be useful in some new type of information processing?\\
$\bullet$ Use of building blocks having multiple physical properties,
each described by a particular group, can cut down resource requirements
by simultaneous execution of multiple transformations.
Such physical objects exist,
e.g. an electron has location, spin, energy level etc.,
and quantum computation has provided the first step in this direction.\\
$\bullet$ The depth of computation can be reduced by direct execution of
complex high level instructions (i.e. without translation to lower levels).
This can be achieved using special purpose components and configurable
systems.
In fact, such features are commonplace in biological systems.
Life started with a nanotechnology---a technology that has now
spanned billions of years and many orders of magnitude.
There is a lot to learn from that regarding hierarchical
execution structures.\\
$\bullet$ Use of large groups can also reduce depth of computation.
Such groups have been used in cryptography, but can we design physical
building blocks that directly implement them?\\
$\bullet$ $(Z_2)^d$ does not provide the minimal set of building blocks
for $d>1$; it contains $2^d$ points compared to $(d+1)$ points
of a simplex---the difference is exponential.
A simplicial geometry can be far more efficient for multi-dimensional
information processing.

Construction of the complete information theory framework
for a general set of building blocks is a wide open subject.
The mathematical definition of information parallels
the thermodynamical definition of entropy.
Entropy just counts the number of available states,
and there is no hurdle in applying it
to configurations of arbitrary building blocks.
The quantification of correlations amongst the building blocks, however,
becomes increasingly complicated as the dimensionality increases.
We have made a start in this direction, in our efforts to incorporate
superposition and entanglement in quantum information theory,
where Boltzmann entropy is generalised to von Neumann entropy.
But there is a long way to go.

Lest we forget, according to the generalised information theory framework
used here, we ourselves are computers.
We will no doubt use whatever devices we can design to augment our
capabilities; humans with gizmos attached to themselves have already
become a reality, and creation of cyborgs would be a natural extension.
But we are also computers that are on the threshold of tinkering with
their own programme, as demonstrated by the recent developments in
molecular biology and genetics.
It is easy to grasp the concept of one computer designing another one,
but a computer that can tune itself and evolve is a different story.
That may make us feel uneasy, even scared at times, but it is inevitable.
Our future is intimately tied to these developments,
and understanding information processing in a wide open framework
would be an inseparable part of it.

\end{document}